\documentclass[10pt,conference]{IEEEtran}
\IEEEoverridecommandlockouts
\usepackage{cite}
\usepackage{amsmath,amssymb,amsfonts}
\usepackage{algorithmic}
\usepackage{graphicx}
\usepackage{textcomp}
\usepackage{xcolor}
\def\BibTeX{{\rm B\kern-.05em{\sc i\kern-.025em b}\kern-.08em
    T\kern-.1667em\lower.7ex\hbox{E}\kern-.125emX}}

\newcommand\toolname{ResTest}
\usepackage{pifont}

\newcommand{\CodeIn}[1]{{\small\texttt{#1}}}
\usepackage[most]{tcolorbox}
\usepackage{multirow}
\usepackage{booktabs}
\usepackage[hidelinks,breaklinks=true]{hyperref}
\usepackage{xurl}
\begin{document}

\title{\toolname: Targeted Coverage of Residual Not-Covered Code Using Large Language Models for Web GUI Testing}

\author{
\IEEEauthorblockN{Chenxu Liu\textsuperscript{1,4,*},
Zhiyu Gu\textsuperscript{3,*},
Guoquan Wu\textsuperscript{3,\ensuremath{\dagger}},
Ying Zhang\textsuperscript{2,4},
Wei Yang\textsuperscript{5},
Jun Wei\textsuperscript{3},
Tao Xie\textsuperscript{1,4,5,\ensuremath{\dagger}}}
\IEEEauthorblockA{\textsuperscript{1}Key Lab of HCST (PKU), MOE; SCS; Peking University, Beijing, China}
\IEEEauthorblockA{\textsuperscript{2}Key Lab of HCST (PKU) \& NERC of SE, MOE; Peking University, Beijing, China}
\IEEEauthorblockA{\textsuperscript{3}Institute of Software Chinese Academy of Sciences, University of Chinese Academy of Sciences, Beijing, China}
\IEEEauthorblockA{\textsuperscript{4}Beijing Tongming Lake Information Technology Application Innovation Center (TLAIC), China}
\IEEEauthorblockA{\textsuperscript{5}Fudan University Institute of Systems for Advanced Computing, China}
\thanks{\textsuperscript{*}Equal contribution.}
\thanks{\textsuperscript{\ensuremath{\dagger}}Corresponding authors.}
}

\maketitle

\begin{abstract}

Automated web GUI testing (AWGT) approaches explore web applications through GUI actions to achieve code coverage. However, existing approaches, whether random-based, model-based, or reinforcement-learning-based, often struggle to generate continuous and semantically meaningful action sequences for testing complex functionalities, limiting their achievable code coverage. Recent LLM-based approaches partially alleviate this problem but still fall short due to limited capability in inferring testable functionalities and low success rates in executing tasks on complex web applications.

In this paper, we propose \toolname, a complementary approach that uses coverage-report-guided LLMs to target residual not-covered code left by existing AWGT approaches. \toolname\ first runs an existing AWGT approach to broadly explore the application while constructing a state transition graph with summarized information. Once coverage plateaus, \toolname\ employs an LLM to infer not-covered functionalities based on the state transition graph and coverage report, and then utilizes a specialized LLM-driven GUI agent to execute these functionalities in a targeted manner.

Our evaluation on ten open-source web applications shows that \toolname\ improves three categories of AWGT approaches by 17.52\% to 24.36\% in average code coverage. Ablation studies further confirm the sustained coverage improvement capability and the contribution of each component.

\end{abstract}


\section{Introduction}
Web applications play an important role across diverse domains, including online marketing, education, and management. To reduce the labor costs associated with quality assurance of web applications, approaches of automated web GUI testing (AWGT) are proposed, aiming to maximize code coverage within a constrained time budget. These approaches interact with the web application under test (AUT) via GUI actions (e.g., clicks, drags), emulating human behavior.

Existing AWGT approaches can be broadly categorized into three main categories.
\textbf{Random-based} approaches~\cite{monkey,visualRandom} generate GUI actions randomly on web pages to perform testing.
\textbf{Model-based} approaches~\cite{crawljax,ATUSA,webembed,fraggen,Judge} dynamically construct a state transition graph using state abstraction techniques~\cite{NDStudy,webembed,Judge} during testing and employ traversal algorithms such as depth-first search to systematically explore the AUT.
\textbf{Reinforcement-Learning-based (RL-based)} approaches~\cite{WebExplor,QExplore,WebQT,UniRLtest,PIRLTest,WebRLED} employ RL algorithms such as Q-Learning~\cite{Q-Learning} to guide the testing process.

However, existing approaches suffer from a limited ability to generate continuous and meaningful action sequences to cover complex functionalities in modern web applications~\cite{VETL,AutoE2E,naviqate}, limiting their ability to reach a satisfying code coverage for two main reasons.
First, existing approaches lack the ability to comprehend content within the AUT, reducing their capacity to complete a single unique functionality in a concentrated manner.
Second, modern web applications are inherently complex, dynamic, and nondeterministic~\cite{crawljax,NDStudy,naviqate}, causing existing approaches to face difficulties in building the state transition graph and exploring these applications.

Recently, a number of studies~\cite{QTypist,InputBlaster,VETL,AutoE2E,naviqate,GPTDroid,DroidAgent,mobileGPT} attempt to leverage the strong semantic understanding and logical reasoning capabilities of large language models (LLMs) to enhance GUI testing, but still face limitations in reaching high code coverage, for three main reasons.
First, testable functionalities are hard to infer. Modern web applications typically exhibit rich, dynamic, and complex content~\cite{VETL}, making it challenging for LLMs to accurately infer functionalities that should be tested, particularly when prioritizing important functionalities (e.g., functionalities that can underly high code coverage) for testing.
Second, the execution context during testing is complex and hard to memorize. A complete testing process can require the generation of hundreds and even thousands of actions, making it challenging for LLMs to recall previously executed actions or already tested functionalities~\cite{VETL,AutoE2E}.
Third, the success rate of task execution is low. Existing work~\cite{mind2web,webarena,visualwebarena,SeeAct,uground} shows that even the state-of-the-art GUI agents achieve limited success rate (i.e., below 30\%) when applied to modern web applications to execute specific tasks. The low success rate inevitably undermines the effectiveness of applying LLMs for AWGT.

To address these limitations, in this paper, we propose \toolname, complementing AWGT with coverage-report-guided LLMs for covering residual not-covered code~\cite{residual} in a targeted manner. \toolname\ consists of four phases.
In the \emph{exploration} phase, \toolname\ runs an existing AWGT approach to broadly explore the AUT until coverage stagnates, while simultaneously collecting GUI information and coverage reports.
In the \emph{asynchronous information processing} phase, \toolname\ organizes the collected information into a state transition graph enriched with natural language descriptions and associations between code files and GUI states.
In the \emph{testing-task inference} phase, \toolname\ employs an LLM to analyze the state transition graph and coverage reports, identifying low-coverage code files and inferring a batch of testing tasks targeting residual not-covered functionalities.
In the \emph{task execution} phase, \toolname\ employs a specialized LLM-driven GUI agent, guided by key-state information from the state transition graph, to execute the inferred tasks and improve coverage.

\toolname\ addresses the aforementioned limitations in the following four ways.
First, to overcome the difficulty of existing AWGT approaches in exercising complex functionalities, \toolname\ introduces an LLM-driven GUI agent that can execute testing tasks in a targeted manner, enabling focused exploration beyond generic strategies.
Second, to mitigate the challenge faced by existing LLM-based AWGT approaches in inferring testable functionalities from complex and information-rich web pages, \toolname\ incorporates coverage reports and an online-constructed state transition graph, and explicitly associates code files with abstracted GUI states. This design allows the LLM to accurately identify not-covered code and reason about corresponding functionalities using contextual information of abstracted GUI states.
Third, to address the limited capability of LLM-based approaches in maintaining complex and prolonged execution contexts, \toolname\ decomposes the overall testing objective into a set of short, self-contained testing tasks, allowing the GUI agent to focus on completing each task within a limited number of action steps.
Fourth, to improve the task success rate of LLM-based AWGT approaches, \toolname\ leverages the state transition graph and key-state information obtained during broad exploration to guide the GUI agent during task execution, thereby increasing execution reliability and effectiveness.

We compare \toolname\ with three widely used AWGT approaches from different categories on ten open-source web applications used in prior work~\cite{DIG,crawljax,WebExplor,WebRLED}. The results show that \toolname\ improves existing approaches by 17.52\% to 24.36\% on average code coverage within a one-hour time budget, demonstrating its effectiveness.
Further evaluations confirm that \toolname\ can continuously identify and cover residual not-covered code, and highlight the effectiveness of incorporating coverage report information. These results demonstrate the effectiveness and general applicability of \toolname.

In summary, this paper makes the following main contributions:
\begin{itemize}
    \item We propose \toolname, the first LLM-enhanced AWGT approach that includes coverage reports to target residual not-covered code, and can be integrated with a wide range of existing AWGT approaches to further improve code coverage.
    \item We design an asynchronous pipeline of information processing that incrementally constructs a state transition graph and enriches it using state descriptions and associations between code files and states, enabling cross-layer reasoning between GUI states and code files.
    \item We present a coverage-guided framework of task inference and execution that utilizes state transition graphs and low-coverage code-file information to infer targeted testing tasks, and employs an LLM-based GUI agent to execute the tasks.
    \item We evaluate \toolname\ on ten widely used open-source web applications. The results show that \toolname\ improves existing approaches by 17.52\% to 24.36\% in code coverage. Our implementation is publicly available~\cite{SeekerRepo}.
\end{itemize}

\section{Approach}
\begin{figure*}[t]
\centerline{\includegraphics[width = \linewidth]{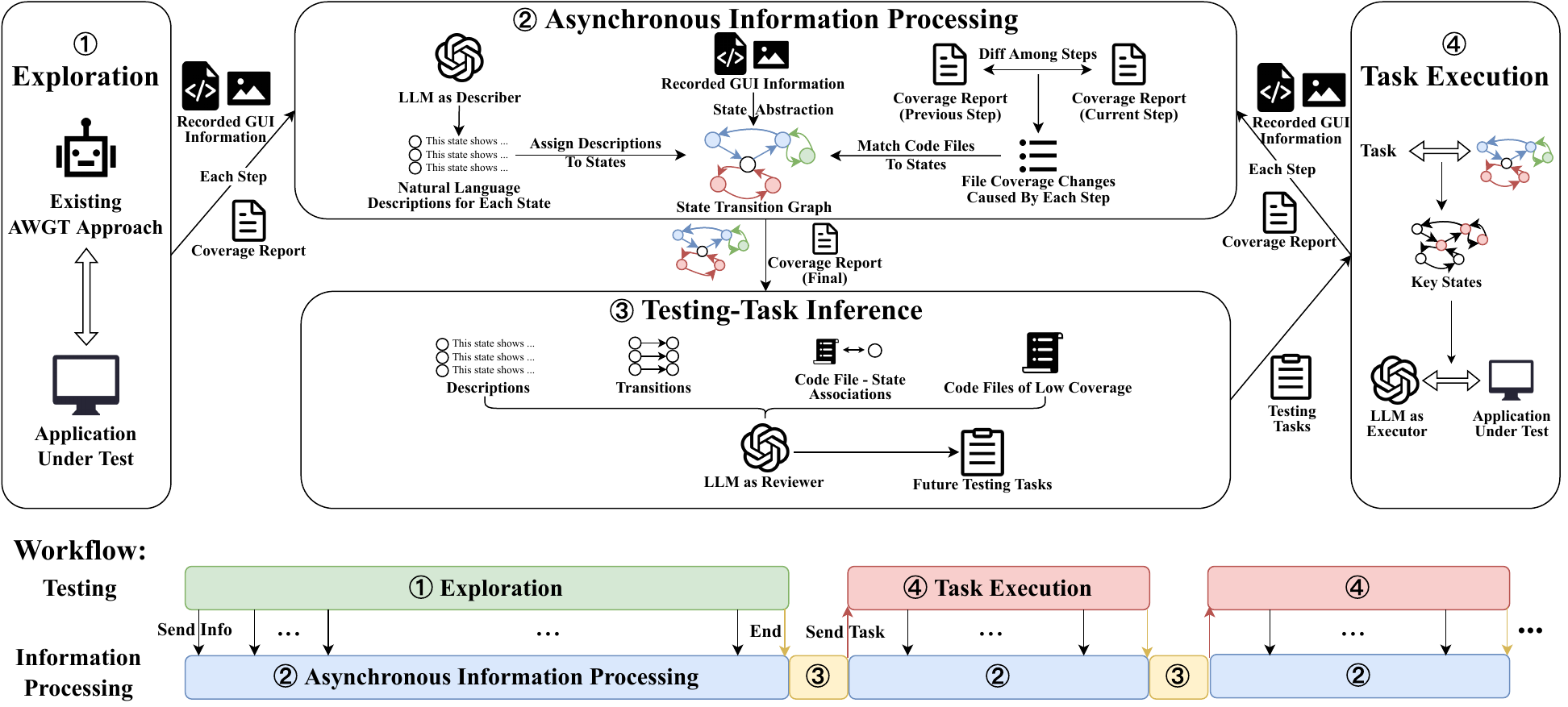}}
\caption{Overview of \toolname.}
\label{structure}
\end{figure*}

Figure~\ref{structure} illustrates the architecture of \toolname, which comprises four phases: exploration (Section~\ref{sec:Exploration}), asynchronous information processing (Section~\ref{sec:InformationProcessing}), testing-task inference (Section~\ref{sec:testing task inference}), and task execution (Section~\ref{MLLMExecution}).


\subsection{Exploration}\label{sec:Exploration}

Because \toolname\ targets the residual not-covered code left by existing AWGT approaches, it first invokes one such approach until code coverage stagnates, thereby broadly exploring the AUT and collecting essential information for subsequent analysis. After each action step, \toolname\ records action-level information (e.g., action type and input value), page-level information (e.g., screenshots and HTML), and the corresponding coverage report. The collected information is forwarded to the asynchronous information processing phase, in which \toolname\ constructs a state transition graph that subsequently guides the LLM in inferring and executing testing tasks targeting residual not-covered code.

\subsection{Asynchronous Information Processing}\label{sec:InformationProcessing}

In this phase, \toolname\ transforms the unstructured, multi-modal information collected during exploration into a concise textual representation stored in a state transition graph, by combining LLM-based analysis with heuristic rules. To avoid blocking the testing process, this phase operates concurrently with the exploration and task execution phases.
The state transition graph is represented by three components: (1) a natural-language description of each state, (2) state transitions, and (3) the code files associated with each state. Templates and instantiations of each component are presented in Table~\ref{tab:template and example}, together with the representation of our coverage report. The resulting textual representation serves as input to the LLM for inferring and executing testing tasks.

\begin{table*}[t]
\caption{Templates and instantiations for each type of processed information used as inputs to the LLMs.}
\begin{center}
\setlength{\tabcolsep}{3pt}
\resizebox{\textwidth}{!}{
\begin{tabular}{p{2cm}|p{4cm}|p{12cm}}
\hline
\textbf{Type} & \textbf{Templates} & \textbf{Instantiations} \\
\hline
Description & \textit{\textless Description\textgreater} &  Description for state 1: The webpage is the account services section of Para Bank, displaying an ``Accounts Overview'' table with account numbers, balances, and available amounts. The user, John Smith, is logged in, and the sidebar offers options like opening a new account, transferring funds, and logging out. The total balance is shown as \$1692.67. Navigation links are available at the top and bottom of the page. \\
\hline
Transition & \textit{\textless Start\textgreater \ + \textless Action\textgreater \ + \textless Value\textgreater \ + \textless Component\textgreater \ + \textless End\textgreater} & Start from State 1; Performed action: click; Action value: ; Performed on component: Transfer Funds link under Account Services; Lead to State 2 \\
\hline
Code File & \{\textit{\textless Filenames\textgreater}\} & \{`com.parasoft.parabank.web.controller/UpdateCustomerController', `com.parasoft.parabank.domain/Cu- stomer', `com.parasoft.parabank.web.controller/RestServiceProx- yController'\} \\
\hline
Coverage Report & \textit{\textless Filename\textgreater \ + \textless Not-covered Lines\textgreater \ + \textless Total Lines\textgreater \ + \textless Not-covered Functions\textgreater} & File name: com.parasoft.parabank.web.controller/UpdateCustomerController;	Missed Lines: 22;	Total Lines: 39;	Not-covered Functions: [`onSubmit', `getCustomer', `setAccessModeController', `setAdminManager']; \\
\hline
\end{tabular}
}
\label{tab:template and example}
\end{center}
\end{table*}

\subsubsection{Construction of the State Transition Graph}
To equip the LLM with sufficient and precise knowledge of the AUT for inferring and executing testing tasks, \toolname\ abstracts the GUI information collected at each action step into a state transition graph that reduces redundancy and complexity.
State transition graphs have been widely adopted in existing AWGT approaches~\cite{NDStudy,TK,webembed,fraggen,Judge}, where pages exhibiting equivalent functionalities from a testing perspective are merged into a single state. Such abstraction mitigates state explosion and avoids redundant exploration of functionally equivalent pages.
We empirically adopt WebEmbed~\cite{webembed} as the state-abstraction technique, since it strikes a favorable balance between abstraction effectiveness and the number of constructed states. After each action step, \toolname\ uses WebEmbed to compare the current page against previously constructed states and either assigns it to an existing state or instantiates a new one.

\subsubsection{Natural Language Descriptions of Each State (Descriptions)}
A constructed state transition graph typically contains dozens of states. Directly representing each state by raw HTML or screenshots would yield an information volume that exceeds the input limits of mainstream LLMs~\cite{GPT2,GPT3}. Moreover, since modern LLMs follow a generative paradigm based on the Transformer architecture~\cite{Transformers}, excessive and redundant information in the prompt impairs reasoning ability and instruction adherence~\cite{LostInTheMiddle}. To convey the knowledge encoded in the state transition graph concisely, \toolname\ generates a natural-language description for each state using a multi-modal LLM (MLLM), referred to as the \textit{Describer}. When a new state is added to the graph, the Describer takes the corresponding screenshot as input and produces a textual description.
Due to space limit, all prompts used in \toolname\ are presented in our open-source repository~\cite{SeekerRepo}.

Because state abstraction is intended to distinguish pages by their testing-relevant functionalities~\cite{NDStudy,TK,webembed,fraggen,Judge}, each description focuses on the functionalities offered by the page and the status of its components.
Since these descriptions also serve the subsequent inference and execution of testing tasks, they should be sufficiently detailed to disambiguate similar states yet concise enough to avoid confusion and redundancy.
The objective is therefore to convert the multi-modal state transition graph into a textual and concise form for use in subsequent phases.
We design the prompt with the Chain-of-Thought (CoT)~\cite{CoT} technique, which guides the LLM to reason step by step before producing the final answer, thereby improving the accuracy and detail of the generated descriptions.

We adopt page screenshots, rather than HTML, as the prompt input for two reasons. First, screenshots preserve the spatial layout of the GUI, allowing the MLLM to leverage its visual understanding to capture component positions and effectively express the visual state of the GUI. Second, since screenshots and prompt text reside in different modalities, the prompt is not flooded by lengthy HTML, reducing the risk of confusing the model or degrading its instruction adherence~\cite{VETL}.


\subsubsection{State Transitions (Transitions)}
After producing the description of each state, \toolname\ also conveys the transitions among states so that the LLM can connect individual states into a coherent view of the entire graph. Because exploration may execute hundreds or even thousands of actions, directly feeding the entire action sequence to the LLM risks both confusion and context-window overflow.
To compactly represent state transitions, \toolname\ first prunes the state transition graph by discarding self-loops (i.e., edges whose source and target are the same state) and retaining a single edge between any pair of states. It then performs a breadth-first traversal over the pruned graph and serializes each transition as a single line of text containing the identifiers of the source and target states, the action type, the action value, and the identifier of the target component.

\subsubsection{Code Files Associated with Each State (Code Files)}
States constructed from GUI information and code coverage reports reside at different abstraction levels and in different modalities. Relying solely on the LLM to bridge them is unreliable. We therefore explicitly associate coverage information with states during the construction of the state transition graph.

After each action step, \toolname\ computes the difference between the current and previous coverage reports and applies two heuristic rules.
First, code files newly observed in the current report are associated with the state reached after the action, since the loading of a new state typically loads its corresponding code files.
Second, code files whose existing coverage changes are associated with the state from which the action was issued, since that action triggered the additional coverage.

To bound time overhead, the differencing is performed at file granularity. Consequently, a code file may be associated with multiple states, and a state may be associated with multiple code files.

\subsubsection{Coverage Report}
Relying solely on the state transition graph to infer functionalities that remain not-covered after the exploration phase is insufficient, because some relevant states may not have been visited or recorded by the AWGT approach, and many functionalities are not visible from GUI pages (e.g., authentication controllers).

To enable effective and targeted generation of testing tasks aimed at residual not-covered code, we augment the LLM context with information extracted from the final coverage report (i.e., the report obtained at the end of the exploration phase or of a previous task execution phase). The coverage report exposes the coverage status of each code file together with the project's organizational structure, helping the LLM identify residual and untested functionalities.

Specifically, \toolname\ ranks code files in the coverage report by the number of not-covered lines and selects the top 25. For each selected file, \toolname\ provides the file path, coverage statistics, and the names of not-covered functions as input to testing-task inference. The coverage report is obtained via code instrumentation, while the names of not-covered functions are extracted via static code analysis.

Because modern web applications are typically built upon mature frameworks and follow standard naming conventions~\cite{styleGuide}, file and function names usually convey their intended functionalities. The coverage report therefore provides the LLM with sufficient and meaningful contextual knowledge to support effective testing-task inference.

\subsection{Testing-Task Inference}
\label{sec:testing task inference}

After the collected information is processed, \toolname\ feeds the resulting concise representation to an LLM, referred to as the \textit{Reviser}, to infer testing tasks targeting functionalities that remain not-covered after the exploration phase. We do not employ an MLLM at this stage, as all relevant information has already been transformed into the textual modality.
The prompt also follows the CoT pattern, instructing the Reviser to first comprehend and summarize each input component before emitting the testing tasks in the specified format. We require the Reviser to output, for each task, the target code file the task aims to cover. This design encourages the Reviser to focus on tasks that are explicitly aligned with improving file-level code coverage.


\subsection{Task Execution}\label{MLLMExecution}
After inferring the testing tasks, \toolname\ executes them concretely to cover residual not-covered functionalities and further improve code coverage. GUI agents~\cite{mind2web,SeeAct,uitars} have recently emerged as an active research topic and are widely applied to task automation. Such agents accept a natural-language task description as input and attempt to complete the task by observing and interacting with the GUI of the target application.

In our task execution phase, we employ a specialized LLM-driven GUI agent, referred to as the \textit{Executor}, to execute testing tasks sequentially. The GUI information observed during execution is fed back to enrich the state transition graph, supporting subsequent rounds of testing-task inference. However, executing testing tasks on real-world web applications remains highly challenging.
Prior studies~\cite{mind2web,SeeAct,uground,aguvis,osatlas,uitars,cogagent,cogvlm,webglm,autowebglm} report that even state-of-the-art GUI agents achieve a success rate below 50\% on realistic benchmarks such as Mind2Web-Live~\cite{mind2web} and WebArena~\cite{webarena}, primarily owing to the inherent complexity of web GUIs and their underlying functionalities.
To improve the task execution success rate, the Executor incorporates two specialized designs: \emph{state-guided navigation} and \emph{planner--actor decoupling}.

\subsubsection{State-Guided Navigation}
Despite the strong generalization capabilities of LLMs, it remains challenging for them to interact effectively with previously unseen web applications~\cite{cogagent,cogvlm}. In \toolname, the state transition graph constructed during testing serves as a shortcut, providing structured domain knowledge of the AUT to support task planning and execution.

However, supplying the entire state transition graph to the Executor at every action step is impractical due to its scale. The excessive amount of information not only impairs effective decision making~\cite{LostInTheMiddle}, but also consumes a large number of input tokens, incurring unnecessary computational overhead. Moreover, only a limited subset of states in the graph is typically relevant to a given task.
For example, an e-commerce system may consist of multiple functional modules (e.g., product listings and shopping carts), leading to a complex state transition graph. In contrast, a task such as modifying the username involves only a few states related to the home page and account settings; including unrelated states does not contribute to effective execution.

To address this issue, before executing each task, \toolname\ identifies the \emph{key states} most relevant to the task by leveraging its target code file. Specifically, for each state in the graph, \toolname\ examines whether the target code file appears in the set of files associated with that state and extracts all states linked to the target file.
\toolname\ then appends the selected key states to the Executor prompt following the template shown in Table~\ref{tab:template and example}. This design supplies the Executor with task-specific contextual knowledge while substantially reducing token consumption and overhead, thereby improving the success rate of task execution.

\subsubsection{Planner-Actor Decoupling}
Inspired by prior work~\cite{SeeAct,uground}, we decouple the action-decision process of the Executor into two modules: a \emph{planner}, which determines the next action, and an \emph{actor}, which locates the corresponding GUI component and performs the action.
Approaches that rely on a single pretrained model for task execution~\cite{cogagent,autowebglm} have limited ability to incorporate external task-specific knowledge (e.g., the key states identified by our approach), since such models are typically trained to consume only the task description.
By contrast, planner--actor decoupling provides greater flexibility and effectiveness in the Executor: it accepts external knowledge, supports customization to the AUT, and enables substitution of components, e.g., interchanging text-based locators~\cite{mind2web}, image-annotation-based locators~\cite{SoM}, and fully vision-based locators~\cite{uground,osatlas,aguvis,uitars} as the actor for GUI component localization.

Concretely, before each step the planner observes the current page and jointly reasons over the task objective, the historical trajectory, and the extracted key states to determine the next action, expressed as a natural-language description. The actor then takes the description as input, locates the target component on the GUI, and performs the action.

The planner prompt (presented in our repository~\cite{SeekerRepo}) comprises a role-play introduction, the available action space, the extracted key states, essential GUI information, the target task, CoT-style analysis guidelines, and a screenshot of the current page.

\section{Implementation}
We implement \toolname\ in Python and deploy it as an asynchronous service using Flask~\cite{Flask}. For state abstraction in the construction of the state transition graph, we adopt WebEmbed~\cite{webembed}, a lightweight technique that balances abstraction effectiveness against the number of constructed states.

We empirically set the time budget of the exploration phase to 30 minutes by default, since prior studies~\cite{WebExplor,WebRLED,Judge} report that the coverage achieved by existing AWGT approaches typically saturates within this duration. This fixed budget also enables consistent analysis of coverage changes before and after the 30-minute mark. Heuristic algorithms~\cite{llmdroid} could alternatively be used to detect saturation and trigger the Executor automatically.

We adopt GPT-4o\footnote{https://platform.openai.com/docs/models/gpt-4o}, a mature and powerful multi-modal LLM, as the backbone model of \toolname\ for processing both textual and visual information. Specifically, GPT-4o serves as the Describer, the Reviser, and the planner of the Executor. For GUI component localization within the Executor, we employ UI-TARS-1.5-7B~\cite{uitars,ui-tars-15-seed} as the actor, which observes the current screenshot and outputs the pixel-level location of the target component. Our Executor is built upon and inspired by SeeAct~\cite{SeeAct}, UGround~\cite{uground}, and WebArena~\cite{webarena}.

The number of testing tasks generated per batch is empirically set to five. Given that \toolname\ completes approximately ten tasks within a 30-minute time budget, this setting balances stable execution against the timely incorporation of newly acquired information, while avoiding redundant testing.

We set the maximum number of execution steps per task to 20, following the convention of WebArena~\cite{webarena}. The planner is also allowed to emit a \CodeIn{[STOP]} action for early termination once it determines that the task has been completed.

\toolname\ is designed to be flexible and generalizable. It does not depend on any specific AWGT approach, state-abstraction technique, MLLM, or planner--actor model, and can be naturally extended with alternative components. \toolname\ can therefore benefit directly from advances in its underlying models and techniques, achieving improved effectiveness when equipped with more capable components.

\section{Evaluation}
Our evaluation is structured around the following five research questions:
\begin{itemize}
    \item \textbf{RQ1 (Effectiveness of \toolname)}: To what extent can \toolname\ enhance the coverage achieved by existing AWGT approaches?
    \item \textbf{RQ2 (Sustained Coverage Improvement)}: To what extent can \toolname\ continuously discover and cover residual code over an extended execution period?
    \item \textbf{RQ3 (Effect of Coverage Report)}: What is the impact of incorporating coverage report information in \toolname\ on its overall effectiveness?
    \item \textbf{RQ4 (Effect of Task Number per Batch)}: How does the number of testing tasks generated by \toolname\ per batch affect its overall effectiveness?
    \item \textbf{RQ5 (Effect of Backbone Model)}: How does the choice of backbone model affect the effectiveness of \toolname?
\end{itemize}

\subsection{Evaluation Setup}

\subsubsection{Baseline Approaches}
To evaluate the effectiveness of \toolname, we compare it against three widely used and state-of-the-art AWGT approaches that represent different categories.
\begin{itemize}
    \item \textbf{Crawljax}~\cite{crawljax}. A model-based AWGT approach widely adopted as a baseline in prior studies~\cite{FeedEx,fraggen,webembed,Judge}. Crawljax employs a depth-first search strategy to explore the AUT and uses string comparison of HTML files for state abstraction.
    \item \textbf{WebExplor}~\cite{WebExplor}. An RL-based AWGT approach that adopts Q-Learning~\cite{Q-Learning} as its exploration strategy and serves as a representative RL-based baseline in prior work~\cite{QExplore,WebQT,WebRLED}.
    \item \textbf{GPTDroid (GPTWeb)}~\cite{GPTDroid}. An LLM-based Android GUI testing approach that uses an LLM to decide each testing action during execution. We migrate it to the web setting and refer to the adapted version as GPTWeb, since existing LLM-based AWGT approaches are either restricted to generating textual inputs and not open-sourced~\cite{VETL}, or focus on testing features rather than code coverage~\cite{AutoE2E,naviqate}.
\end{itemize}

For Crawljax and WebExplor, we use their publicly available implementations with the configurations recommended by their authors. We extend both tools by inserting network requests to our \toolname\ service after each action step and upon exhausting the exploration time budget.

Since GPTDroid is originally designed for Android and cannot be directly applied to web applications, we adapt it to the web setting through three targeted modifications.
First, we align its action space with the WebArena framework~\cite{webarena}.
Second, we revise its prompts (e.g., replacing Android activity names with URLs) and incorporate additional contextual information from the prompt template of UGround~\cite{uground} to enrich its knowledge.
Third, we apply our planner--actor decoupling design and replace its HTML/XML-based state representation with screenshots, leveraging the strengths of MLLMs to mitigate the long-context issue~\cite{LostInTheMiddle} caused by verbose web HTML~\cite{VETL}.

We release our adapted version, GPTWeb, in our open-source repository~\cite{SeekerRepo} for reproducibility. To ensure a fair comparison, we use identical login scripts across all evaluated approaches and a fixed interval of 1,000 ms between actions.

According to existing work~\cite{crawljax,fraggen,NDStudy,FeedEx,WebExplor,WebQT,QExplore,WebRLED,Judge}, AWGT approaches exhibit randomness during testing. To mitigate this randomness, we repeat each experiment five times and report the average results, following common practice~\cite{WebQT,WebRLED}. We allocate a one-hour time budget for all experiments. Specifically, for \toolname, we allocate 30 minutes to exploration and 30 minutes to task execution, since prior work has shown that 30 minutes is sufficient for existing AWGT approaches to reach coverage saturation~\cite{Judge,WebQT,WebExplor}.

\subsubsection{Application Subjects}
We adopt ten open-source web applications widely used in prior studies~\cite{fraggen,WebExplor,WebQT,webembed,WebRLED} as the evaluation subjects. Detailed information about these applications is provided in Table~\ref{tab:webapps_info}.

The applications are divided into two groups.
The first group, shown on the left side of the table, was originally introduced by DIG~\cite{DIG}. These applications were collected in 2019 and instrumented using the JavaScript instrumentation tool NYC~\cite{NYC}, with branch coverage as the coverage metric.
The second group, shown on the right side, was introduced by a recent study, WebRLED~\cite{WebRLED}. These applications were collected in 2025 and instrumented using NYC~\cite{NYC} and JaCoCo~\cite{Jacoco} for server-side coverage, with line coverage as the coverage metric.

Considering both the collection time and the complexity of the applications, we refer to the two groups as \emph{simple} and \emph{complex}, respectively, following the criterion of Do{\u{g}}an et al.~\cite{dougan2014web}, under which applications with more than 10,000 lines of code (LOC) are regarded as complex.
We exclude Petclinic~\cite{Petclinic} from the simple group because it contains only 20 branches and can be fully covered through straightforward navigation and form filling, leaving little room to differentiate the evaluated approaches. We exclude 4gaBoards~\cite{4gaBoards} from the complex group because \toolname\ consistently triggers an inherent fault that crashes the server-side service; restarting the service resets coverage, preventing reliable evaluation. Notably, this observation also suggests the potential of \toolname\ in exposing latent faults.

\begin{table*}[t]
    \centering
    \caption{Detailed information of our application subjects.}
    \setlength{\tabcolsep}{2pt}
    \begin{tabular}{c|cccc||c|cccc}
        \hline
        \textbf{Name}   & \textbf{Version} & \textbf{Client LOC} & \textbf{Server LOC} & \textbf{Domain} & \textbf{Name}   & \textbf{Version} & \textbf{Client LOC} & \textbf{Server LOC} & \textbf{Domain} \\
        \hline
        Dimeshift~\cite{Dimeshift} & 2018 & 5,140 (JS) & 3,298 (JS) & Finance & Agilefant~\cite{agilenfant}  & 3.5.4 & 3,949 (JSP) & 27,584 (Java) &  Management \\

        Pagekit~\cite{Pagekit} & 1.0.15 & 4,214 (JS) & 13,856 (PHP) & Publishing & Gadael~\cite{gadael} & 0.1.4 & 6,265 (JS) & 7,811 (Java) & Management \\

        Phoenix~\cite{Phoenix} & 1.1 & 2,289 (JS) & 1,135 (Elixir) & Management & Parabank~\cite{parabank}  & 2024 & 2,662 (JSP) & 9,446 (Java) & Finance \\

        Retroboard~\cite{Retroboard} & 2019 & 2,144 (JS) & 278 (JS) & Collaboration & Realworld~\cite{realworld} & 2024 & 7,604 (JS) & 2,705 (TS) & Blog  \\

        Splittypie~\cite{Splittypie} & 2018 & 2,710 (JS) & 829 (JS) & Finance &  Timeoff~\cite{timeoff} & 1.0.0 & 2,937 (Handlebars) & 7,933 (JS) & Attendance \\
        \hline
    \end{tabular}

    \label{tab:webapps_info}
\end{table*}


\subsubsection{Evaluation Environment}
We conduct our evaluation on an Ubuntu 22.04 PC equipped with an Intel i9-10900K CPU and 32 GB of RAM. The actor model of the Executor is deployed as a Flask~\cite{Flask} service on a remote server equipped with an NVIDIA RTX 4090 GPU and accessed over the network.

\subsection{RQ1: Effectiveness of \toolname}
\begin{figure*}[t]
    \centering
    \includegraphics[width=0.85\linewidth]{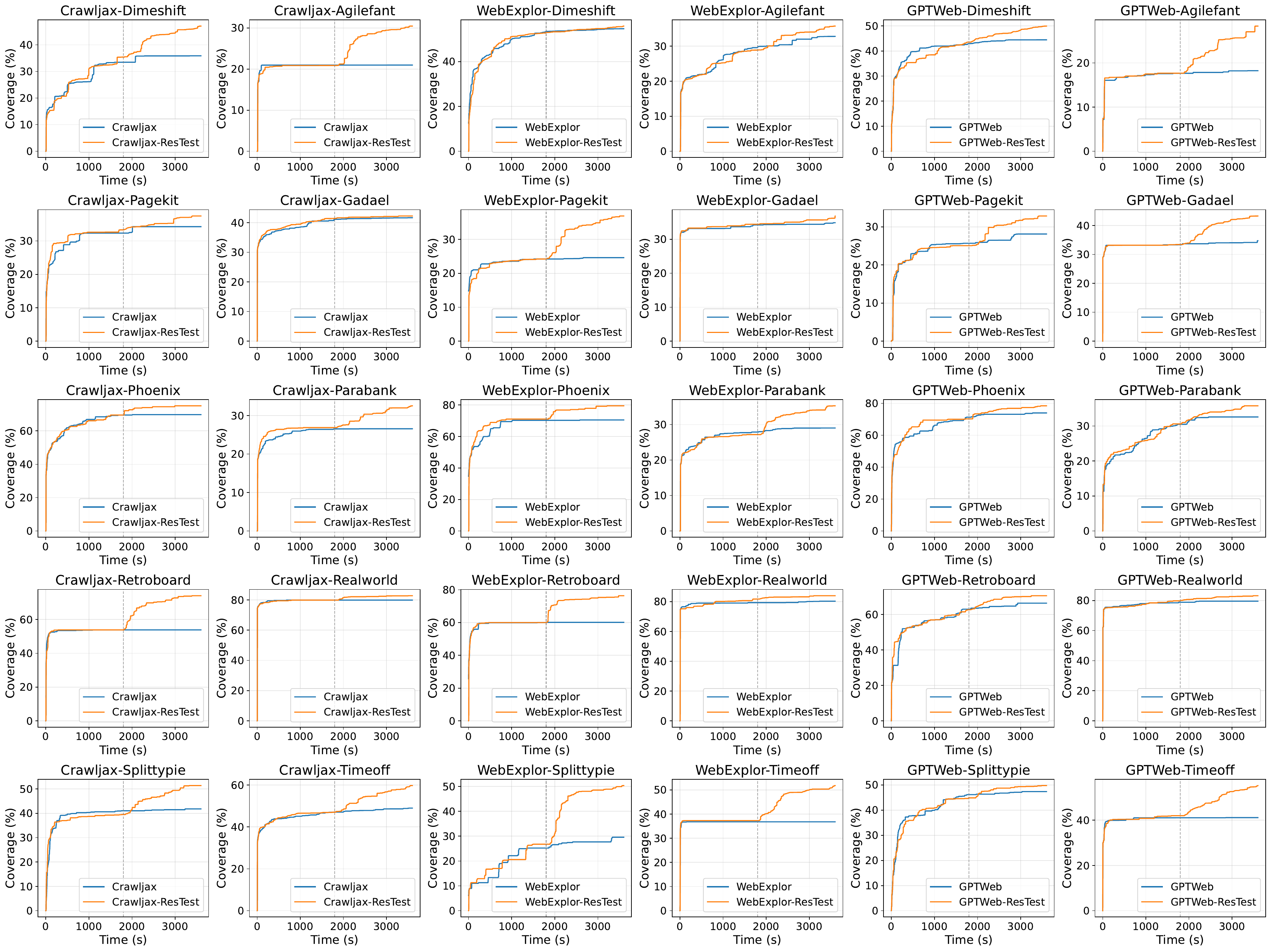}
    \caption{Coverage trends over time for baseline approaches with and without \toolname.}
    \label{RQ1Figure}
\end{figure*}

We evaluate the effectiveness of \toolname\ by comparing each of the three baseline AWGT approaches with its augmented version that integrates \toolname. The achieved code coverage is reported in Table~\ref{RQ1Coverage}, the coverage trends over time are illustrated in Figure~\ref{RQ1Figure}, and the number of distinct failures triggered by each approach is reported in Table~\ref{RQ1Failures}.

\begin{table*}[t]
    \centering
    \begin{minipage}[t]{0.74\textwidth}
    \centering
    \caption{Code coverage results of \toolname\ and baseline approaches.}
    \setlength{\tabcolsep}{2pt}
    \resizebox{\textwidth}{!}{
\begin{tabular}{@{}c|ccc|ccc|ccc@{}}
\toprule
           & \multicolumn{3}{c|}{Crawljax}                                         & \multicolumn{3}{c|}{WebExplor}                                        & \multicolumn{3}{c}{GPTWeb}                                           \\ \midrule
App        & w/o \toolname & w/ \toolname & Improv. & w/o \toolname & w/ \toolname & Improv. & w/o \toolname & w/ \toolname & Improv. \\ \midrule
Dimeshift  & 35.89\%                     & 47.06\%                      & +31.12\% & 54.67\%                     & 55.83\%                      & +2.12\%  & 44.44\%                     & 49.91\%                      & +12.31\% \\
Pagekit    & 34.27\%                     & 37.46\%                      & +9.31\%  & 24.60\%                     & 36.89\%                      & +49.96\% & 28.13\%                     & 32.86\%                      & +16.81\% \\
Phoenix    & 69.74\%                     & 75.00\%                      & +7.54\%  & 70.53\%                     & 79.47\%                      & +12.68\% & 73.95\%                     & 78.42\%                      & +6.04\%  \\
Retroboard & 53.80\%                     & 74.04\%                      & +37.62\% & 60.12\%                     & 76.37\%                      & +27.03\% & 66.43\%                     & 70.64\%                      & +6.34\%  \\
Splittypie & 41.79\%                     & 51.37\%                      & +22.92\% & 29.57\%                     & 50.34\%                      & +70.24\% & 47.35\%                     & 49.74\%                      & +5.05\%  \\ \midrule
Agilefant  & 20.97\%                     & 30.45\%                      & +45.21\% & 32.77\%                     & 35.71\%                      & +8.97\%  & 18.25\%                     & 28.34\%                      & +55.29\% \\
Gadael     & 41.67\%                     & 42.26\%                      & +1.42\%  & 34.88\%                     & 36.87\%                      & +5.71\%  & 34.76\%                     & 43.31\%                      & +24.60\% \\
Parabank   & 26.57\%                     & 32.53\%                      & +22.43\% & 29.02\%                     & 35.27\%                      & +21.54\% & 32.57\%                     & 35.75\%                      & +9.76\%  \\
Realworld  & 79.87\%                     & 82.77\%                      & +3.63\%  & 80.29\%                     & 83.98\%                      & +4.60\%  & 79.56\%                     & 83.20\%                      & +4.58\%  \\
Timeoff    & 49.01\%                     & 59.75\%                      & +21.91\% & 36.78\%                     & 51.79\%                      & +40.81\% & 41.14\%                     & 55.30\%                      & +34.42\% \\ \midrule
Average    & 45.36\%                     & 53.27\%                      & +20.31\% & 45.32\%                     & 54.25\%                      & +24.36\% & 46.66\%                     & 52.75\%                      & +17.52\% \\ \bottomrule
\end{tabular}}

    \label{RQ1Coverage}
    \end{minipage}
    \hfill
    \begin{minipage}[t]{0.245\textwidth}
    \centering
    \caption{Number of distinct failures triggered by \toolname\ and baseline approaches.}
    \setlength{\tabcolsep}{2pt}
    \resizebox{\textwidth}{!}{
\begin{tabular}{@{}c|cc|cc|cc@{}}
\toprule
           & \multicolumn{2}{c|}{Crawljax} & \multicolumn{2}{c|}{WebExplor} & \multicolumn{2}{c}{GPTWeb} \\ \midrule
App        & w/o  & w/  & w/o  & w/  & w/o & w/  \\ \midrule
Dimeshift  & 8    & 9   & 8    & 12  & 5   & 5   \\
Pagekit    & 8    & 26  & 9    & 17  & 5   & 10  \\
Phoenix    & 7    & 9   & 3    & 6   & 2   & 1   \\
Retroboard & 2    & 2   & 1    & 2   & 1   & 1   \\
Splittypie & 7    & 6   & 6    & 7   & 6   & 3   \\ \midrule
Agilefant  & 4    & 6   & 10   & 10  & 2   & 2   \\
Gadael     & 11   & 9   & 9    & 8   & 6   & 9   \\
Parabank   & 3    & 6   & 5    & 9   & 5   & 8   \\
Realworld  & 8    & 9   & 10   & 11  & 4   & 6   \\
Timeoff    & 1    & 2   & 7    & 8   & 1   & 2   \\ \midrule
Average    & 5.9  & 8.4 & 6.8  & 9   & 3.7 & 4.7 \\ \bottomrule
\end{tabular}}
    \label{RQ1Failures}
    \end{minipage}
\end{table*}

Table~\ref{RQ1Coverage} is organized into three blocks, one per baseline. Within each block, the three columns report (i) the coverage of the original baseline after one hour of testing (\textit{w/o \toolname}); (ii) the coverage of the same baseline augmented with \toolname\ (\textit{w/ \toolname}); and (iii) the relative improvement, computed as $(c_r - c_b) / c_b \times 100\%$, where $c_b$ and $c_r$ denote the coverage of the baseline and of the augmented version, respectively.

The results show that \toolname\ consistently outperforms all three baselines across all ten applications. Compared with Crawljax, WebExplor, and GPTWeb, \toolname\ yields average coverage improvements of 20.31\%, 24.36\%, and 17.52\%, respectively. These results strongly demonstrate the effectiveness of \toolname\ in covering the residual not-covered code left by existing AWGT approaches. Moreover, the consistent gains across all applications indicate that \toolname\ generalizes well and is not tailored to specific application characteristics or to a particular category of baseline.

As shown in Figure~\ref{RQ1Figure}, the curves of \toolname\ (orange) clearly dominate those of the baselines (blue). A vertical dashed line in each subplot marks the 30-minute mark, after which \toolname\ activates its LLM-driven GUI agent.
Before the 30-minute mark, \toolname\ achieves coverage comparable to its baseline, since the asynchronous information processing in \toolname\ does not interfere with the baseline's exploration. After this mark, the coverage curve of \toolname\ continues to rise sharply, whereas the curves of the baselines plateau. This contrast further evidences the ability of \toolname\ to cover residual not-covered code left by existing approaches.

Even after averaging over multiple runs, minor coverage differences may remain at the 30-minute mark due to the inherent randomness of AWGT approaches. Such differences do not affect our conclusions for three reasons.
First, when \toolname\ shows slightly higher coverage at 30 minutes, the coverage of the baseline still does not exceed its own subsequent plateau.
Second, when a baseline shows slightly higher coverage at 30 minutes, \toolname\ overtakes it within the next 30 minutes.
Third, the final coverage gap between \toolname\ and the baselines is substantially larger than any randomness-induced variation observed at the 30-minute mark.

For some applications (e.g., Gadael and Realworld), \toolname\ provides only limited additional coverage. This is attributable to factors such as unreachable code and the inherent limitations of current LLM capabilities, which we discuss in detail in Section~\ref{sec:covnot100}.

Beyond code coverage, we additionally evaluate the effectiveness of \toolname\ in terms of the number of distinct failures triggered during testing. Following prior work~\cite{Judge,WebQT,WebExplor,WebRLED}, we treat system-level errors reported in the browser's console as failures and manually deduplicate the triggered failures within each run, removing failures that differ only in parameter values, in line with the procedure adopted by WebRLED~\cite{WebRLED}.

The number of distinct failures triggered by each approach is reported in Table~\ref{RQ1Failures}. Compared with Crawljax, WebExplor, and GPTWeb, \toolname\ raises the average failures triggered by 42.37\%, 32.35\%, and 27.03\%, respectively. The most pronounced gains are observed on Pagekit and Parabank, where the targeted execution of low-coverage code paths exposes a substantially larger set of latent failures.

We also observe that, on a small number of applications and baselines, \toolname\ triggers slightly fewer failures than the original baseline. This result is because \toolname\ reallocates part of the testing budget from the baseline's broad exploration to targeted task execution focused on covering specific code files. On applications where the baseline's exploration happens to traverse many failure-prone paths, this reallocation may modestly reduce the incidental discovery of failures. Such regressions are small in absolute terms and are outweighed on average by the gains elsewhere. Since \toolname\ is not explicitly designed for failure finding,  its consistent average improvement across all three baselines indicates that the broader and more targeted code paths it exercises also surface additional latent failures as a useful by-product of improving code coverage.

\subsection{RQ2: Sustained Coverage Improvement}

We assess the capability of \toolname\ to continuously analyze and cover residual not-covered code by comparing it with a variant equipped with a longer time budget. The variant runs the baseline approach for 30 minutes, followed by 90 minutes of LLM-based task inference and execution, for a total budget of two hours. We denote this variant by \toolname$_{2h}$ and report its results in Table~\ref{RQ2Coverage}.

For this and the remaining RQs, we adopt only GPTWeb as the baseline. This choice is motivated by two considerations: GPTWeb achieves the highest average coverage among the three baselines on all ten applications, and, as an LLM-based approach, it provides the most appropriate baseline for evaluating the advantages of \toolname\ over LLM-based approaches.
We further restrict the experimental scope to the complex group, since coverage on the simple group already approaches saturation within one hour, leaving little room to differentiate \toolname\ from its variants.

\begin{table}[t]
    \centering
    \caption{Code coverage results of \toolname\ and ablation versions on the complex group of application subjects, with GPTWeb as the baseline approach.}
    \setlength{\tabcolsep}{2pt}
    \begin{tabular}{@{}c|c|c|c|cc@{}}
    \toprule
    App        & \toolname & \toolname$_{2h}$ & \toolname$_{NoCov}$ & \toolname$_{TN2}$ & \toolname$_{TN10}$  \\ \midrule
    Agilefant  &   28.34\%                       &    36.70\%                      &    24.59\%      &   25.81\%                                &         24.65\%                                                          \\
    Gadael     &   43.31\%                       &     47.51\%                     &     42.61\%             &    42.51\%                               &        46.31\%                                                   \\
    Parabank   &   35.75\%                       &     38.06\%                    &    31.94\%    &   35.05\%                                  &     35.50\%                                                               \\
    Realworld  &   83.20\%                       &      83.46\%                    &      82.81\%       &    81.89\%                             &      82.44\%                                                            \\
    Timeoff    &    55.30\%                      &    63.89\%                      &     53.14\%     &    56.61\%                                     &      54.12\%                                                       \\ \midrule
    Average    &    49.18\%                      &     53.92\%                      &      47.02\%       &    48.37\%                                 &       48.60\%                                                      \\ \bottomrule
    \end{tabular}
    \label{RQ2Coverage}
\end{table}

\toolname$_{2h}$ consistently outperforms the one-hour version of \toolname\ across all five applications, with an average relative improvement of 9.64\%. This result demonstrates the strong capability of \toolname\ to continuously increase coverage over time.
Benefiting from continuous information processing and iterative updates of both the state transition graph and the coverage report, \toolname\ continues to gain coverage throughout the two-hour run. In contrast, the baseline approaches converge within 30 minutes to one hour, as shown in Figure~\ref{RQ1Figure}. This contrast further highlights the effectiveness of \toolname\ in sustaining coverage improvement beyond the saturation point of existing approaches.


\subsection{RQ3: Effect of Coverage Report}
We conduct an ablation study on the coverage report incorporated into the state transition graph and the LLM prompts to examine its contribution to inferring and executing testing tasks. Specifically, we remove the \CodeIn{[Code File]} and \CodeIn{[Coverage Report]} components from the Reviser prompt, and remove the related-state information from the Executor. The results of this ablation are reported in Table~\ref{RQ2Coverage} under the column \toolname$_{NoCov}$.

\toolname\ consistently outperforms \toolname$_{NoCov}$ across all five applications, with an average relative improvement of 4.59\%. This result indicates that the state transition graph alone is insufficient for effectively inferring not-covered functionalities.
By exposing the code files mapped to each state and identifying low-coverage files, the coverage report provides direct and fine-grained guidance toward unexercised functionalities. Equipped with this enriched signal, the LLM can prioritize tasks that are more likely to exercise not-covered code, even when such tasks lie far from the states already visited in the state transition graph.


\subsection{RQ4: Effect of Task Number per Batch}
We investigate how the number of testing tasks generated per batch during the testing-task inference phase affects \toolname. By default, \toolname\ generates five testing tasks per batch. We evaluate two additional variants that generate two and ten tasks per batch, denoted by \toolname$_{TN2}$ and \toolname$_{TN10}$ in Table~\ref{RQ2Coverage}.

\toolname$_{TN10}$ achieves the highest coverage on Gadael, while \toolname$_{TN2}$ performs best on Timeoff. The default version of \toolname\ achieves the highest coverage on the remaining three applications. These results suggest that the optimal batch size is application-dependent: some applications require more frequent knowledge updates to flexibly identify not-covered functionalities, whereas others benefit from concentrating on a larger set of inferred tasks within a single batch and avoiding the disruption introduced by overly frequent iterations.

Notably, the default version is never the worst configuration on any application and achieves the highest average coverage among all variants. This indicates that a batch size of five strikes a favorable balance between adaptability and execution focus. Overall, the differences across all variants remain within 1\% on average, showing that \toolname\ is not sensitive to this parameter.

\subsection{RQ5: Effect of Backbone Model}
We examine the impact of the backbone model on the effectiveness of \toolname\ by replacing the default GPT-4o with alternative open-source MLLMs~\cite{qwen3}. Specifically, \toolname$_{Q235B}$ replaces GPT-4o with Qwen3-VL-235B-A22B-Instruct, and \toolname$_{Q30B}$ replaces it with Qwen3-VL-30B-A3B-Instruct. The replacement applies to the Describer, the Reviser, and the planner of the Executor, while the actor remains UI-TARS-1.5-7B. The results are presented in Table~\ref{ChangeModelCoverage}.

All variants of \toolname\ consistently outperform the GPTWeb baseline across all five applications. Compared with GPTWeb, the default version of \toolname\ (with GPT-4o) achieves an average relative improvement of 19.19\%, while \toolname$_{Q30B}$ and \toolname$_{Q235B}$ achieve 15.58\% and 7.56\%, respectively. Even when the proprietary backbone is replaced with smaller open-source models, \toolname\ still delivers substantial coverage gains, indicating that its effectiveness stems from the approach itself rather than from a particular backbone model.

Notably, \toolname$_{Q30B}$ outperforms \toolname$_{Q235B}$ overall and even surpasses the default version on Timeoff. This observation suggests that effectiveness is not solely determined by parameter scale: reasoning and instruction-following abilities are equally critical for \toolname. By manually inspecting the execution logs, we find that Qwen-235B exhibits weak instruction adherence as the planner, frequently violating the ``one GUI action per step'' instruction by emitting multiple (sometimes dozens of) actions per turn. Since our implementation executes only the first returned action, this inflates the reasoning cost and wastes tokens and time, ultimately yielding lower coverage.

The default version achieves the highest average coverage, indicating that more capable backbone models can further enhance \toolname. This observation highlights the orthogonal design between \toolname\ and its backbone model. Owing to budget constraints, we do not evaluate more advanced models such as GPT-5.5~\cite{GPT-5-5} or Claude Opus 4.8~\cite{Claude-opus-4-8}. As the reasoning capabilities of both open-source and proprietary MLLMs continue to advance, \toolname\ is expected to benefit directly without requiring methodological changes.

\begin{table}[t]
    \centering
    \caption{Code coverage results of \toolname\ and variants with different backbone models on the complex group of application subjects, with GPTWeb as the baseline approach.}
    \begin{tabular}{@{}ccccc@{}}
    \toprule
    App        & GPTWeb & \toolname &  \toolname$_{Q235B}$ &  \toolname$_{Q30B}$ \\ \midrule
    Agilefant  &  18.25\% &  28.34\%                       &    23.25\%        &    25.16\%                                      \\
    Gadael     &  34.76\%  & 43.31\%                       &     39.63\%       &    40.05\%                                        \\
    Parabank   &  32.57\%  & 35.75\%                       &     33.44\%       &    35.48\%                                        \\
    Realworld  &  79.56\%  & 83.20\%                       &      80.94\%      &    81.00\%                                        \\
    Timeoff    &  41.14\%   & 55.30\%                      &    44.66\%        &    56.75\%                                            \\ \midrule
    Average    &   41.26\%  & 49.18\%                      &     44.38\%           &    47.69\%                                    \\ \bottomrule
    \end{tabular}
    \label{ChangeModelCoverage}
\end{table}


\section{Threats to Validity}

\textbf{Internal Validity.}
First, the configurations of both the baselines and \toolname\ may affect the results. We mitigate this by adopting the settings recommended in prior work~\cite{WebExplor,crawljax} for the baselines and configuring all approaches with identical login scripts and time budget.
Second, randomness may threaten internal validity. We repeat each experiment five times and report the average, following common practice~\cite{WebRLED,Judge,WebQT,WebExplor,QExplore}. In Figure~\ref{RQ1Figure}, the curves largely overlap during the first 30 minutes, reflecting the stability of our experiments.
Third, faults may exist in our implementation, which we mitigate through pair programming and systematic code reviews.

\textbf{External Validity.}
The selection of application subjects may introduce bias. We mitigate this by adopting representative open-source web applications widely used in prior studies~\cite{DIG,WebRLED}, which exhibit sufficient complexity and are representative of real-world AWGT scenarios.

\section{Discussion}

\subsection{Final Coverage Achieved by \toolname}
\label{sec:covnot100}
According to Table~\ref{RQ2Coverage}, the coverage on some applications remains below 50\% even after two hours of testing. This arises from two factors.

First, GUI testing and code instrumentation impose inherent ceilings on the attainable coverage. GUI interactions cannot reach code paths unrelated to UI behavior, such as backend data processing (e.g., parsing uploaded files), error handling, background tasks, and unreachable legacy code; instrumentation tools may further include third-party libraries. To validate this, we manually tested Parabank for 30 minutes and obtained only 43\% coverage. Since human testers operate more than ten times faster than LLM-driven agents, this result indicates an inherent coverage ceiling.

Second, current LLM capabilities impose practical constraints. We observe cases in which the planner repeatedly interacts with non-interactable components or falls into loops. As \toolname\ is orthogonal to its underlying LLMs, it should benefit directly from future advances in LLM reasoning.

\subsection{Cost of \toolname}
Compared with non-LLM AWGT approaches, \toolname\ inevitably introduces additional monetary cost, which is justified by its substantial coverage gains.
Empirically, 30 minutes of task inference and execution costs approximately \$0.8 with GPT-4o, plus a negligible overhead (less than \$0.02) during the first 30 minutes for state-description generation, since WebEmbed constructs only a limited number of states. As shown in RQ5, testers can further reduce cost by replacing the backbone with open-source MLLMs such as Qwen3-VL~\cite{qwen3}, achieving competitive effectiveness at a fraction of the cost.

\subsection{Comparison with Mobile Testing}
LLM-based mobile GUI testing approaches~\cite{QTypist,InputBlaster,GPTDroid,DroidAgent,Trident,llmdroid} differ fundamentally from \toolname\ and face three challenges when applied to web applications.
First, web applications exhibit richer and more dynamic contexts than mobile applications~\cite{VETL} (e.g., complex layouts, asynchronous content loading, and diverse interaction patterns), so prompts and strategies tailored to mobile environments tend to lose effectiveness in web settings.
Second, most mobile testing approaches target the main functionalities of the AUT rather than fine-grained code coverage, whereas \toolname\ explicitly reasons about not-covered code and iteratively guides testing toward coverage improvement.
Third, LLM-based agents achieve substantially lower task-execution and component-localization success rates on web applications than on mobile platforms~\cite{uground,cogagent}, owing to the increased complexity and variability of web GUIs.
To enable a fair comparison, we migrated GPTDroid to the web setting, yielding GPTWeb. Our results show that \toolname\ further improves its coverage by targeting and covering the residual not-covered code.


\section{Related Work}

\subsection{Automated Web GUI Testing}
Automated web GUI testing has been investigated by a wide range of approaches, e.g., TESTAR~\cite{testar,testar1}, GUITAR~\cite{guitar}, Crawljax~\cite{crawljax}, ATUSA~\cite{ATUSA}, FeedEx~\cite{FeedEx}, White et al.~\cite{visualRandom}, NDStudy~\cite{NDStudy}, Corazza et al.~\cite{TK}, FragGen~\cite{fraggen}, WebEmbed~\cite{webembed}, Judge~\cite{Judge}, WebExplor~\cite{WebExplor}, WebQT~\cite{WebQT}, QExplor~\cite{QExplore}, UniRLTest~\cite{UniRLtest}, PIRLTest~\cite{PIRLTest}, and WebRLED~\cite{WebRLED}. Requiring neither a pre-constructed model of the AUT nor its source code, these black-box approaches explore the AUT under random-based~\cite{monkey,visualRandom}, model-based~\cite{guitar,testar1,crawljax,ATUSA,FeedEx,NDStudy,fraggen,webembed,Judge}, or RL-based~\cite{WebExplor,WebQT,QExplore,WebRLED} strategies, often combined with state abstraction~\cite{NDStudy,TK,fraggen,webembed} to avoid redundant exploration. However, they struggle to generate continuous and meaningful action sequences for testing complex functionalities. \toolname\ addresses this limitation by integrating the semantic understanding and reasoning capabilities of LLMs into AWGT.

\subsection{LLMs for GUI Testing}
With the rapid progress of LLMs, recent approaches exploit their capabilities for GUI testing. QTypist~\cite{QTypist}, InputBlaster~\cite{InputBlaster}, and VETL~\cite{VETL} use LLMs to generate meaningful textual inputs. GPTDroid~\cite{GPTDroid}, DroidAgent~\cite{DroidAgent}, MobileGPT~\cite{mobileGPT}, MemoDroid~\cite{memodroid}, and Trident~\cite{DroidAgent} test the main functionalities of the AUT through question answering with an LLM. Guardian~\cite{guardian}, AutoE2E~\cite{AutoE2E}, and NaviQAte~\cite{naviqate} infer the main functionalities of the AUT or seek to raise the success rate of executing them. LLMDroid~\cite{llmdroid} uses an LLM to help existing approaches escape exploration tarpits~\cite{vet} when coverage stagnates. However, these approaches are constrained by the limited ability of LLMs to comprehend the AUT and to execute tasks reliably, resulting in limited coverage. \toolname\ addresses these limitations by incorporating coverage report information for targeted testing and leveraging the constructed state transition graph to guide task execution.

\subsection{GUI Agent}
Mind2Web~\cite{mind2web} introduces the concept of a generalist GUI agent that executes natural-language tasks on arbitrary web applications. A series of LLM-based agents~\cite{cogagent,cogvlm,webglm,autowebglm} have been trained to improve task-execution success on web applications. SeeAct~\cite{SeeAct} first divides a GUI agent into a planning phase and a grounding phase, relieving LLMs from complex single-step decision making, after which subsequent work~\cite{uground,aguvis,osatlas,uitars} focuses on training vision-based grounding models to improve grounding accuracy. Nevertheless, recent studies~\cite{mind2web,webarena,visualwebarena,workarena,workarena++} report that existing GUI agents still suffer from low task-execution success rates. \toolname\ exploits the constructed state transition graph as task-specific knowledge to improve the success rate of task planning and execution, covering complex functionalities of the AUT in a targeted manner.

\section{Conclusion}
In this paper, we propose \toolname, the first coverage-report-guided approach that targets the residual not-covered code left by existing automated approaches of web GUI testing.
Motivated by the limited coverage achieved by existing approaches, \toolname\ jointly leverages an online-constructed state transition graph and coverage reports, and exploits the reasoning capabilities of LLMs to infer and execute testing tasks in a targeted manner.
We design a novel LLM-based mechanism that summarizes multi-modal information, identifies not-covered functionalities, and executes the corresponding testing tasks, with prompts crafted using the Chain-of-Thought technique.
Our evaluation shows that \toolname\ substantially improves the code coverage of widely used AWGT approaches by 17.52\% to 24.36\%, demonstrating its effectiveness and practical applicability.

\section{Data Availability}
All data, results, and source code of ResTest are available in our Zenodo repository~\cite{SeekerRepo}.

\bibliographystyle{IEEEtran}
\bibliography{sample-base}

@misc{monkey,
  title={Monkey},
  url="https://developer.android.com",
  year="2022"
}

@misc{GPT-5-5,
  title =       {{Introducing GPT-5.5}},
  year =         {2026},
  url =          {https://openai.com/index/introducing-gpt-5-5/},
}

@misc{Claude-opus-4-8,
  title =       {{Introducing Claude Opus 4.8}},
  year =         {2026},
  url =          {https://www.anthropic.com/news/claude-opus-4-8},
}

@misc{Splittypie,
  title =       "Splittypie",
  year =         {2025},
  url =          {https://github.com/tsubik/splittypie},
}

@misc{Pagekit,
  title =       {Pagekit},
  year =         {2023},
  url =          {https://github.com/pagekit/pagekit},
}

@misc{Petclinic,
  title =       {{Spring Petclinic Angular}},
  year =         {2025},
  url =          {https://github.com/spring-petclinic/spring-petclinic-angular},
}

@misc{Phoenix,
  title =       {{Phoenix Trello}},
  year =         {2025},
  url =          {https://github.com/bigardone/phoenix-trello},
}

@misc{Dimeshift,
  title =       {Dimeshift},
  year =         {2025},
  url =          {https://github.com/jeka-kiselyov/dimeshift},
}

@misc{Retroboard,
  title =       "Retroboard",
  year =         {2025},
  url =          {https://github.com/antoinejaussoin/retro-board},
}

@misc{timeoff,
  title={{Timeoff Management Application}},
  year="2023",
  url="https://github.com/timeoff-management/timeoff-management-application",
}

@misc{4gaBoards,
  title={{4ga Boards}},
  year="2025",
    url="https://github.com/RARgames/4gaBoards",
}

@misc{gadael,
  title="Gadael",
  year="2020",
  url="https://github.com/gadael/gadael",
}

@misc{parabank,
  title="Parabank",
  year="2024",
  url="https://github.com/parasoft/parabank",
}

@misc{realworld,
  title="Realworld",
  year="2022",
  url="https://github.com/gothinkster/realworld",
}

@misc{agilenfant,
  title="Agilenfant",
  year="2016",
  url="https://sourceforge.net/projects/agilefant",
}

@misc{styleGuide,
  title={{Google Style Guide}},
  year="2025",
  url="https://google.github.io/styleguide",
}

@misc{SeekerRepo,
	title={{Replication Package and Evaluation Results of ResTest}},
	url={https://zenodo.org/records/20509828},
	year={2026}
}

@misc{Jacoco,
  title={{JaCoCo - Java Code Coverage}},
  year="2025",
  url="https://www.eclemma.org/jacoco/",
}

@misc{NYC,
  title={{NYC/Istanbul, a JavaScript test coverage tool}},
  year="2025",
  url="https://istanbul.js.org/",
}

@misc{ui-tars-15-seed,
  title = {{ByteDance Seed. UI-TARS-1.5}},
  year="2025",
  url="https://seed-tars.com/1.5",
}

@book{Flask,
  author       = {Miguel Grinberg},
  title  =  {Flask web development: {Developing} web applications with {Python}},
  year         = {2014}
}

@inproceedings{DIG,
  author       = {Matteo Biagiola and
                  Andrea Stocco and
                  Filippo Ricca and
                  Paolo Tonella},
  title        = {Diversity-based web test generation},
  booktitle    = {Proceedings of the 2019 {ACM} Joint Meeting on European Software Engineering Conference and Symposium on the Foundations of Software Engineering},
  year         = {2019},
}

@inproceedings{vet,
  title={Vet: {I}dentifying and avoiding {UI} exploration tarpits},
  author={Wang, Wenyu and Yang, Wei and Xu, Tianyin and Xie, Tao},
  booktitle={Proceedings of the 29th ACM Joint Meeting on European Software Engineering Conference and Symposium on the Foundations of Software Engineering},
  year={2021}
}

@inproceedings{residual,
  title={Residual test coverage monitoring},
  author={Pavlopoulou, Christina and Young, Michal},
  booktitle={Proceedings of the 21st International Conference on Software Engineering},
  year={1999}
}

@article{qwen3,
  title={{Qwen3-VL} technical report},
  author={Bai, Shuai and Cai, Yuxuan and Chen, Ruizhe and Chen, Keqin and Chen, Xionghui and Cheng, Zesen and Deng, Lianghao and Ding, Wei and Gao, Chang and Ge, Chunjiang and others},
  journal={arXiv preprint arXiv:2511.21631},
  year={2025}
}

@article{SoM,
  title={Set-of-mark prompting unleashes extraordinary visual grounding in {GPT-4V}},
  author={Yang, Jianwei and Zhang, Hao and Li, Feng and Zou, Xueyan and Li, Chunyuan and Gao, Jianfeng},
  journal={arXiv preprint arXiv:2310.11441},
  year={2023}
}

@article{webembed,
  title={Neural embeddings for web testing},
  author={Stocco, Andrea and Willi, Alexandra and Starace, Luigi Libero Lucio and Biagiola, Matteo and Tonella, Paolo},
  journal={arXiv preprint arXiv:2306.07400},
  year={2023}
}

@article{WebRLED,
  title={Deep reinforcement learning for automated web {GUI} testing},
  author={Gu, Zhiyu and Liu, Chenxu and Wu, Guoquan and Zhang, Yifei and Yang, ChenXi and Liang, Zheheng and Chen, Wei and Wei, Jun},
  journal={arXiv preprint arXiv:2504.19237},
  year={2025}
}

@article{naviqate,
  title={{NaviQAte}: {Functionality-guided} web application navigation},
  author={Shahbandeh, Mobina and Alian, Parsa and Nashid, Noor and Mesbah, Ali},
  journal={arXiv preprint arXiv:2409.10741},
  year={2024}
}

@article{DroidAgent,
  title={Autonomous large language model agents enabling intent-driven mobile {GUI} testing},
  author={Yoon, Juyeon and Feldt, Robert and Yoo, Shin},
  journal={arXiv preprint arXiv:2311.08649},
  year={2023}
}

@article{testar,
  title={{TESTAR}: {Tool} support for test automation at the user interface level},
  author={Vos, Tanja EJ and Kruse, Peter M and Condori-Fern{\'a}ndez, Nelly and Bauersfeld, Sebastian and Wegener, Joachim},
  journal={International Journal of Information System Modeling and Design (IJISMD)},
  year={2015}
}

@article{testar1,
  title={{TESTAR}--{S}criptless testing through graphical user interface},
  author={Vos, Tanja EJ and Aho, Pekka and Pastor Ricos, Fernando and Rodriguez-Valdes, Olivia and Mulders, Ad},
  journal={Software Testing, Verification and Reliability},
  year={2021}
}

@article{guitar,
  title={{GUITAR}: an innovative tool for automated testing of {GUI}-driven software},
  author={Nguyen, Bao N and Robbins, Bryan and Banerjee, Ishan and Memon, Atif},
  journal={Automated software engineering},
  year={2014}
}

@article{Trident,
  title={Vision-driven automated mobile {GUI} testing via multimodal large language model},
  author={Liu, Zhe and Li, Cheng and Chen, Chunyang and Wang, Junjie and Wu, Boyu and Wang, Yawen and Hu, Jun and Wang, Qing},
  journal={arXiv preprint arXiv:2407.03037},
  year={2024}
}

@article{uground,
  title={Navigating the digital world as humans do: Universal visual grounding for {GUI} agents},
  author={Gou, Boyu and Wang, Ruohan and Zheng, Boyuan and Xie, Yanan and Chang, Cheng and Shu, Yiheng and Sun, Huan and Su, Yu},
  journal={arXiv preprint arXiv:2410.05243},
  year={2024}
}

@article{osatlas,
  title={{OS-ATLAS}: A foundation action model for generalist {GUI} agents},
  author={Wu, Zhiyong and Wu, Zhenyu and Xu, Fangzhi and Wang, Yian and Sun, Qiushi and Jia, Chengyou and Cheng, Kanzhi and Ding, Zichen and Chen, Liheng and Liang, Paul Pu and others},
  journal={arXiv preprint arXiv:2410.23218},
  year={2024}
}

@article{aguvis,
  title={Aguvis: Unified pure vision agents for autonomous {GUI} interaction},
  author={Xu, Yiheng and Wang, Zekun and Wang, Junli and Lu, Dunjie and Xie, Tianbao and Saha, Amrita and Sahoo, Doyen and Yu, Tao and Xiong, Caiming},
  journal={arXiv preprint arXiv:2412.04454},
  year={2024}
}

@article{uitars,
  title={{UI-TARS}: {Pioneering} automated {GUI} interaction with native agents},
  author={Qin, Yujia and Ye, Yining and Fang, Junjie and Wang, Haoming and Liang, Shihao and Tian, Shizuo and Zhang, Junda and Li, Jiahao and Li, Yunxin and Huang, Shijue and others},
  journal={arXiv preprint arXiv:2501.12326},
  year={2025}
}

@article{webarena,
  title={{WebArena}: {A} realistic web environment for building autonomous agents},
  author={Zhou, Shuyan and Xu, Frank F and Zhu, Hao and Zhou, Xuhui and Lo, Robert and Sridhar, Abishek and Cheng, Xianyi and Ou, Tianyue and Bisk, Yonatan and Fried, Daniel and others},
  journal={arXiv preprint arXiv:2307.13854},
  year={2023}
}

@article{visualwebarena,
  title={{VisualWebArena}: Evaluating multimodal agents on realistic visual web tasks},
  author={Koh, Jing Yu and Lo, Robert and Jang, Lawrence and Duvvur, Vikram and Lim, Ming Chong and Huang, Po-Yu and Neubig, Graham and Zhou, Shuyan and Salakhutdinov, Ruslan and Fried, Daniel},
  journal={arXiv preprint arXiv:2401.13649},
  year={2024}
}

@article{workarena,
  title={{WorkArena}: How capable are web agents at solving common knowledge work tasks?},
  author={Drouin, Alexandre and Gasse, Maxime and Caccia, Massimo and Laradji, Issam H and Del Verme, Manuel and Marty, Tom and Boisvert, L{\'e}o and Thakkar, Megh and Cappart, Quentin and Vazquez, David and others},
  journal={arXiv preprint arXiv:2403.07718},
  year={2024}
}

@article{llmdroid,
  title={{LLMDroid}: Enhancing automated mobile app {GUI} testing coverage with large language model guidance},
  author={Wang, Chenxu and Liu, Tianming and Zhao, Yanjie and Yang, Minghui and Wang, Haoyu},
  journal={Proceedings of the ACM on Software Engineering},
  year={2025}
}

@inproceedings{GPT3,
  author       = {Tom B. Brown and
                  Benjamin Mann and
                  Nick Ryder and
                  Melanie Subbiah and
                  Jared Kaplan and
                  Prafulla Dhariwal and
                  Arvind Neelakantan and
                  Pranav Shyam and
                  Girish Sastry and
                  Amanda Askell and
                  Sandhini Agarwal and
                  Ariel Herbert{-}Voss and
                  Gretchen Krueger and
                  Tom Henighan and
                  Rewon Child and
                  Aditya Ramesh and
                  Daniel M. Ziegler and
                  Jeffrey Wu and
                  Clemens Winter and
                  Christopher Hesse and
                  Mark Chen and
                  Eric Sigler and
                  Mateusz Litwin and
                  Scott Gray and
                  Benjamin Chess and
                  Jack Clark and
                  Christopher Berner and
                  Sam McCandlish and
                  Alec Radford and
                  Ilya Sutskever and
                  Dario Amodei},
  title        = {Language models are few-shot learners},
  booktitle    = {Proceedings of the 34th International Conference on Neural Information Processing Systems},
  year         = {2020}
}

@inproceedings{guardian,
  title={Guardian: A runtime framework for {LLM}-based {UI} exploration},
  author={Ran, Dezhi and Wang, Hao and Song, Zihe and Wu, Mengzhou and Cao, Yuan and Zhang, Ying and Yang, Wei and Xie, Tao},
  booktitle={Proceedings of the 33rd International Symposium on Software Testing and Analysis},
  year={2024}
}

@inproceedings{visualRandom,
  title={Improving random {GUI} testing with image-based widget detection},
  author={White, Thomas D and Fraser, Gordon and Brown, Guy J},
  booktitle={Proceedings of the 28th ACM SIGSOFT International Symposium on Software Testing and Analysis},
  year={2019}
}

@inproceedings{NDStudy,
  title={Near-duplicate detection in web app model inference},
  author={Yandrapally, Rahulkrishna and Stocco, Andrea and Mesbah, Ali},
  booktitle={Proceedings of the 42nd international conference on software engineering},
  year={2020}
}

@inproceedings{TK,
  title={Web application testing: Using tree kernels to detect near-duplicate states in automated model inference},
  author={Corazza, Anna and Di Martino, Sergio and Peron, Adriano and Starace, Luigi Libero Lucio},
  booktitle={Proceedings of the 15th International Symposium on Empirical Software Engineering and Measurement},
  year={2021}
}

@inproceedings{WebExplor,
  title={Automatic web testing using curiosity-driven reinforcement learning},
  author={Zheng, Yan and Liu, Yi and Xie, Xiaofei and Liu, Yepang and Ma, Lei and Hao, Jianye and Liu, Yang},
  booktitle={Proceedings of the 43rd International Conference on Software Engineering},
  year={2021}
}

@inproceedings{WebQT,
  title={A reinforcement learning approach to generating test cases for web applications},
  author={Chang, Xiaoning and Liang, Zheheng and Zhang, Yifei and Cui, Lei and Long, Zhenyue and Wu, Guoquan and Gao, Yu and Chen, Wei and Wei, Jun and Huang, Tao},
  booktitle={Proceedings of the 2023 International Conference on Automation of Software Test},
  year={2023}
}

@inproceedings{mobileGPT,
  title={{MobileGPT}: Augmenting {LLM} with human-like app memory for mobile task automation},
  author={Lee, Sunjae and Choi, Junyoung and Lee, Jungjae and Wasi, Munim Hasan and Choi, Hojun and Ko, Steve and Oh, Sangeun and Shin, Insik},
  booktitle={Proceedings of the 30th Annual International Conference on Mobile Computing and Networking},
  year={2024}
}

@inproceedings{UniRLtest,
  title={{UniRLTest}: Universal platform-independent testing with reinforcement learning via image understanding},
  author={Zhang, Ziqian and Liu, Yulei and Yu, Shengcheng and Li, Xin and Yun, Yexiao and Fang, Chunrong and Chen, Zhenyu},
  booktitle={Proceedings of the 31st International Symposium on Software Testing and Analysis},
  year={2022}
}

@inproceedings{memodroid,
  title={Beyond static {GUI} agent: {E}volving {LLM}-based {GUI} testing via dynamic memory},
  author={Chen, Mengzhuo and Liu, Zhe and Chen, Chunyang and Wang, Junjie and Xue, Yangguang and Wu, Boyu and Huang, Yuekai and Wu, Libin and Wang, Qing},
  booktitle={Proceedings of the 40th IEEE/ACM International Conference on Automated Software Engineering},
  year={2025},
  organization={IEEE}
}

@inproceedings{AutoE2E,
  title={Feature-driven end-to-end test generation},
  author={Alian, Parsa and Nashid, Noor and Shahbandeh, Mobina and Shabani, Taha and Mesbah, Ali},
  booktitle={Proceedings of the 47th International Conference on Software Engineering},
  year={2025}
}

@inproceedings{autowebglm,
  title={{AutoWebGLM}: A large language model-based web navigating agent},
  author={Lai, Hanyu and Liu, Xiao and Iong, Iat Long and Yao, Shuntian and Chen, Yuxuan and Shen, Pengbo and Yu, Hao and Zhang, Hanchen and Zhang, Xiaohan and Dong, Yuxiao and others},
  booktitle={Proceedings of the 30th ACM SIGKDD Conference on Knowledge Discovery and Data Mining},
  year={2024}
}

@inproceedings{cogagent,
  title={Cogagent: A visual language model for {GUI} agents},
  author={Hong, Wenyi and Wang, Weihan and Lv, Qingsong and Xu, Jiazheng and Yu, Wenmeng and Ji, Junhui and Wang, Yan and Wang, Zihan and Dong, Yuxiao and Ding, Ming and others},
  booktitle={Proceedings of the IEEE/CVF Conference on Computer Vision and Pattern Recognition},
  year={2024}
}

@inproceedings{SeeAct,
  author       = {Boyuan Zheng and
                  Boyu Gou and
                  Jihyung Kil and
                  Huan Sun and
                  Yu Su},
  title        = {{GPT-4V}(ision) is a generalist web agent, if grounded},
  booktitle    = {Proceedings of the 41th International Conference on Machine Learning},
  year         = {2024}
}

@inproceedings{FeedEx,
  author       = {Amin Milani Fard and
                  Ali Mesbah},
  title        = {Feedback-directed exploration of web applications to derive test models},
  booktitle    = {Proceedings of the 24th International Symposium on Software Reliability Engineering},
  year         = {2013}
}

@inproceedings{QTypist,
  title={Fill in the blank: Context-aware automated text input generation for mobile {GUI} testing},
  author={Liu, Zhe and Chen, Chunyang and Wang, Junjie and Che, Xing and Huang, Yuekai and Hu, Jun and Wang, Qing},
  booktitle={Proceedings of the 45th International Conference on Software Engineering},
  year={2023}
}

@inproceedings{InputBlaster,
  title={Testing the limits: Unusual text inputs generation for mobile app crash detection with large language model},
  author={Liu, Zhe and Chen, Chunyang and Wang, Junjie and Chen, Mengzhuo and Wu, Boyu and Tian, Zhilin and Huang, Yuekai and Hu, Jun and Wang, Qing},
  booktitle={Proceedings of the 46th International Conference on Software Engineering},
  year={2024}
}

@inproceedings{VETL,
  title={Leveraging large vision-language model for better automatic web {GUI} testing},
  author={Wang, Siyi and Wang, Sinan and Fan, Yujia and Li, Xiaolei and Liu, Yepang},
  booktitle={Proceedings of the 2024 IEEE International Conference on Software Maintenance and Evolution},
  year={2024}
}

@inproceedings{GPTDroid,
  title={Make {LLM} a testing expert: Bringing human-like interaction to mobile {GUI} testing via functionality-aware decisions},
  author={Liu, Zhe and Chen, Chunyang and Wang, Junjie and Chen, Mengzhuo and Wu, Boyu and Che, Xing and Wang, Dandan and Wang, Qing},
  booktitle={Proceedings of the 46th International Conference on Software Engineering},
  year={2024}
}

@inproceedings{Transformers,
  author       = {Ashish Vaswani and
                  Noam Shazeer and
                  Niki Parmar and
                  Jakob Uszkoreit and
                  Llion Jones and
                  Aidan N. Gomez and
                  Lukasz Kaiser and
                  Illia Polosukhin},
  title        = {Attention is all you need},
  booktitle    = {Proceedings of the 31st International Conference on Neural Information Processing Systems},
  year         = {2017},
}

@inproceedings{webglm,
  author       = {Xiao Liu and
                  Hanyu Lai and
                  Hao Yu and
                  Yifan Xu and
                  Aohan Zeng and
                  Zhengxiao Du and
                  Peng Zhang and
                  Yuxiao Dong and
                  Jie Tang},
  title        = {{WebGLM}: Towards an efficient and reliable web-enhanced question answering system},
  booktitle    = {Proceedings of the 29th {ACM} {SIGKDD} Conference on Knowledge Discovery
                  and Data Mining},
  year         = {2023}
}

@article{dougan2014web,
  title={Web application testing: A systematic literature review},
  author={Do{\u{g}}an, Serdar and Betin-Can, Aysu and Garousi, Vahid},
  journal={Journal of Systems and Software},
  year={2014}
}

@article{Judge,
author = {Liu, Chenxu and Wang, Junheng and Yang, Wei and Zhang, Ying and Xie, Tao},
title = {Judge: Effective state abstraction for guiding automated web {GUI} testing},
year = {2025},
journal={ACM Transactions on Software Engineering and Methodology},
}

@article{CoT,
  title={Chain-of-thought prompting elicits reasoning in large language models},
  author={Wei, Jason and Wang, Xuezhi and Schuurmans, Dale and Bosma, Maarten and Xia, Fei and Chi, Ed and Le, Quoc V and Zhou, Denny and others},
  journal={Advances in Neural Information Processing Systems},
  year={2022}
}

@article{GPT2,
  title={Language models are unsupervised multitask learners},
  author={Radford, Alec and Wu, Jeffrey and Child, Rewon and Luan, David and Amodei, Dario and Sutskever, Ilya and others},
  journal={OpenAI blog},
  year={2019}
}

@article{Q-Learning,
  title={Q-learning},
  author={Watkins, Christopher JCH and Dayan, Peter},
  journal={Machine Learning},
  year={1992}
}

@article{workarena++,
  title={{WorkArena++}: Towards compositional planning and reasoning-based common knowledge work tasks},
  author={Boisvert, L{\'e}o and Thakkar, Megh and Gasse, Maxime and Caccia, Massimo and de Chezelles, Thibault and Cappart, Quentin and Chapados, Nicolas and Lacoste, Alexandre and Drouin, Alexandre},
  journal={Advances in Neural Information Processing Systems},
  year={2024}
}

@article{cogvlm,
  title={{CogVLM}: Visual expert for pretrained language models},
  author={Wang, Weihan and Lv, Qingsong and Yu, Wenmeng and Hong, Wenyi and Qi, Ji and Wang, Yan and Ji, Junhui and Yang, Zhuoyi and Zhao, Lei and XiXuan, Song and others},
  journal={Advances in Neural Information Processing Systems},
  year={2024}
}

@article{ATUSA,
  title={Invariant-based automatic testing of modern web applications},
  author={Mesbah, Ali and Van Deursen, Arie and Roest, Danny},
  journal={IEEE Transactions on Software Engineering},
  year={2011}
}

@article{crawljax,
  title={Crawling {Ajax}-based web applications through dynamic analysis of user interface state changes},
  author={Mesbah, Ali and Van Deursen, Arie and Lenselink, Stefan},
  journal={ACM Transactions on the Web (TWEB)},
  year={2012}
}

@article{fraggen,
  title={Fragment-based test generation for web apps},
  author={Yandrapally, Rahul Krishna and Mesbah, Ali},
  journal={IEEE Transactions on Software Engineering},
  year={2022}
}

@article{mind2web,
  title={{Mind2Web}: Towards a generalist agent for the web},
  author={Deng, Xiang and Gu, Yu and Zheng, Boyuan and Chen, Shijie and Stevens, Sam and Wang, Boshi and Sun, Huan and Su, Yu},
  journal={Advances in Neural Information Processing Systems},
  year={2023}
}

@article{PIRLTest,
  title={Effective, platform-independent {GUI} testing via image embedding and reinforcement learning},
  author={Yu, Shengcheng and Fang, Chunrong and Li, Xin and Ling, Yuchen and Chen, Zhenyu and Su, Zhendong},
  journal={ACM Transactions on Software Engineering and Methodology},
  year={2024}
}

@article{QExplore,
  title={{QExplore}: An exploration strategy for dynamic web applications using guided search},
  author={Sherin, Salman and Muqeet, Asmar and Khan, Muhammad Uzair and Iqbal, Muhammad Zohaib},
  journal={Journal of Systems and Software},
  year={2023}
}

@article{LostInTheMiddle,
  title={Lost in the middle: {How} language models use long contexts},
  author={Liu, Nelson F and Lin, Kevin and Hewitt, John and Paranjape, Ashwin and Bevilacqua, Michele and Petroni, Fabio and Liang, Percy},
  journal={Transactions of the Association for Computational Linguistics},
  year={2024}
}

\end{document}